\documentclass{aa}
\usepackage{graphicx}
\usepackage{natbib}
\bibpunct{(}{)}{;}{a}{}{,}

\begin{document}

\title{Trapping of the \ion{H}{II} and Photodissociation
Region in a Radially Stratified Molecular Cloud}

\author{Takashi Hosokawa}

\offprints{\email{hosokawa@th.nao.ac.jp}}
\institute{Division of Theoretical Astrophysics, 
National Astronomical Observatory, Mitaka, Tokyo 181-8588, Japan}

\abstract
%contect
{}
%aims
{We study the expansion of the ionization and dissociation fronts (DFs)
in a radially stratified molecular cloud, whose density distribution
is represented as $n(r) \propto r^{-w}$. We focus on cases
with $w \leq 1.5$, when the ionization front is ``trapped'' in the
cloud and expands with the preceding shock front.
The simultaneous evolution of the outer photodissociation region
(PDR) is examined in detail. }
%methods
{ First, we analytically probe the time evolution of the column
densities of the shell and envelope outside the \ion{H}{II} region, 
which are key physical
quantities for the shielding of dissociating photons.
Next, we perform numerical calculations, and study
how the thermal/chemical structure of the outer PDR
changes with different density gradients. 
We apply our numerical model to the 
Galactic \ion{H}{II} region, Sharpless 219 (Sh219). }
%results
{ The time evolution of the column densities of the shell and outer
envelope depends on $w$, and qualitatively changes across $w = 1$.
In the cloud with $w < 1$, the shell column density increases
as the \ion{H}{II} region expands. The DFs are finally trapped in the shell,
and the molecular gas gradually accumulates in the shell.
The molecular shell and envelope surround the \ion{H}{II} region.
With $w > 1$, on the other hand, the shell column density initially
increases, but finally decreases. The column density of the outer
envelope also quickly decreases as the \ion{H}{II} region swells up.
It becomes easier and easier for the dissociating photons to
penetrate the shell and envelope. The PDR
broadly extends around the trapped \ion{H}{II} region. A model with
$w = 1.5$ successfully explains the observational properties
of Sh219. Our model suggests that a density-bounded
PDR surrounds the photon-bounded \ion{H}{II} region in Sh219. 
}
%conclusions
{}


\titlerunning{Trapping of \ion{H}{II} region and PDR}
\authorrunning{Hosokawa}

\maketitle

%%%%%%%%%%%%%%%%%%%%%%%
\section{Introduction}
%%%%%%%%%%%%%%%%%%%%%%%

\ion{H}{II} regions are one of the basic elements of the 
interstellar medium and have been studied by many authors.
Roughly speaking, the theoretical modeling has been advanced
from two complementary standpoints.
Some studies have modeled the detailed thermal and chemical 
structure of the static \ion{H}{II} region \citep[e.g.,][]{Fr98}
and photodissociation region \citep[PDR, e.g.,][]{TH85,Ab05}.
We can compare these models with the observed 
line-strengths and their ratios to probe the physical state of
the target region. Other studies have focused on the
dynamical time evolution of the expanding \ion{H}{II} region 
\citep[e.g.,][]{Yk86}.
The density and temperature actually evolve as the \ion{H}{II} region 
expands, and the gas flows across the ionization front (IF).
A shock front (SF) sometimes precedes the IF
owing to high pressure in the \ion{H}{II} region.
The dynamical evolution is different in different density 
distributions \citep[][ hereafter FTB90]{FTB90}.
The ionized gas sometimes blows out from the molecular cloud
as a ``champagne flow'' \citep{Tt79}.

Confronting some recent observations, however, we require up-to-date 
modeling which includes both the fluid dynamics and consistent
thermal/chemical structure \citep[e.g.,][]{Fr03, Hn06}.
For example, the ``collect and collapse'' by the expanding 
\ion{H}{II} region is one of the dynamical triggering processes of   
star formation \citep[e.g.,][]{EL77, El98}.
Recently, high-resolution observations have revealed the detailed
structure of the shell around some pc-scale \ion{H}{II} regions 
\citep{Deh03a, Deh05, Zv06}.  The dense shell-like structures are 
traced by the molecular and dust emission just around the ionized gas.
Young clusters including OB stars are often embedded in 
fragments of the shell, which is along the ``collect and collapse''
scenario. In our previous papers 
\citep[][ hereafter, Papers I and II]{HI05, HI06}, 
we have focused on the fact that a cold molecular 
rather than warm neutral shell surrounds the \ion{H}{II} region. 
We have analyzed the dynamical 
expansion of the \ion{H}{II} region, PDR, and the swept-up shell, solving
the UV$(h\nu > 13.6~{\rm eV})$- and FUV$(h\nu < 13.6~{\rm eV})$-radiative 
transfer, thermal and chemical
processes in a time-dependent hydrodynamics code.
Our numerical calculations have shown excellent agreement with 
some observational properties of a "collect and collapse" candidate.
The PDR is quickly trapped in the shell, and a cold and 
dense molecular shell forms 
around the \ion{H}{II} region in a {\it homogeneous} ambient medium.

In this paper, we examine how such basic time evolution changes
in an {\it inhomogeneous} ambient medium. Since massive stars form
 in a dense region in the molecular cloud, such a situation 
will be more realistic than expansion in the homogeneous medium.
Actually, observational features of some \ion{H}{II} regions 
are very different from those of the ``collect
and collapse'' candidates. 
For example, the Galactic \ion{H}{II} region, 
Sharpless 219 (Sh219) has no dense shell. 
Only a diffuse neutral layer surrounds the ionized gas \citep{RL93}, 
though the sign of triggering is shown in the
adjacent molecular cloud \citep{Deh03a,Deh06}.
We show that these observational properties are successfully
explained by our model, where the \ion{H}{II} region and PDR expand
in a radially stratified molecular cloud.

The structure of this paper is as follows:
In \S~\ref{sec:ana}, we briefly probe the expansion of the \ion{H}{II}
region and PDR in the radial density gradient with an analytic 
treatment. We focus on the time evolution
of the column densities of the shell and envelope, 
which are key physical quantities for the shielding of FUV radiation. 
In \S~\ref{sec:calc}, we investigate the dynamical evolution and
trapping of the IF and DFs in the molecular cloud
using detailed numerical calculations.
We apply our numerical modeling to Sh219 in \S~\ref{sec:sh219}. 
We assign \S~\ref{sec:dis} and \ref{sec:conc} to discussions and
conclusions.

%%%%%%%%%%%%%%%%%%%%%%%%%%%%
\section{Analytic Treatment}
\label{sec:ana}
%%%%%%%%%%%%%%%%%%%%%%%%%%%%%

In this paper, we study the simultaneous expansion
of the \ion{H}{II} region and surrounding PDR in an 
ambient medium presenting a radial density gradient. 
We adopt a density distribution 
including a central core and outer envelope, 
\begin{equation}
n(r) = \left\{
         \begin{array}{ll}
          n_{\rm c}, & \quad \mbox{for $r < R_{\rm c}$} \\
          \displaystyle{
          n_{\rm c} \left( \frac{r}{R_{\rm c}} \right)^{-w}
          },
          & \quad \mbox{for $r > R_{\rm c}$ }
         \end{array}\right.
\label{eq:nr}
\end{equation}
where $R_{\rm c}$ is the core radius, $n_{\rm c}$ is the core 
density, and $w (>0)$ is the power-law index of the density profile
in the outer envelope (FTB90). First, we consider the expansion
of the \ion{H}{II} region in the analytic approach, following FTB90.
The expansion is subject to the balance between the
recombination rate in the \ion{H}{II} region and supply rate of 
UV photons by the star. Unless this balance is achieved, 
the IF quickly propagates in the molecular cloud as an R-type front.
It is not until the IF runs out of the cloud that the
ionized gas expands owing to the pressure gradient 
(radial ``champagne flow'').
 Once the UV photon supply is balanced by the recombination, 
on the other hand, the IF decelerates and is ``trapped'' in the cloud.
The IF changes from R-type to D-type, and the SF emerges in 
front of the IF. FTB90 have shown that this can occur when $w \leq 1.5$.
At the moment that the ionization balance is achieved, the radius of 
the \ion{H}{II} region is, 
\begin{equation}
R_w  = R_{\rm st}
      \left( \frac{R_{\rm st}}{R_{\rm c}} \right)^{2w/(3-2w)} 
      \left[
       \frac{3 - 2w}{3} + \frac{2 w}{3} 
                          \left( \frac{R_{\rm c}}{R_{\rm st}} \right)
      \right]^{1/(3 - 2w)} 
\end{equation}
for $w < 1.5$, and
\begin{equation}
R_{1.5} =  R_{\rm c} \exp \left[
                  \frac13 \left[ 
                  \left( \frac{R_{\rm st}}{R_{\rm c}} \right)^3
                  - 1     \right]
                \right]
\end{equation}
for $w = 1.5$, where $R_{\rm st}$ is the Str\"omgren radius 
corresponding to the core density.
Note that $R_w$ becomes equal to $R_{\rm st}$ with $w = 0$. 
The subsequent expansion is driven by the high pressure
of the \ion{H}{II} region.
A shell forms around the \ion{H}{II} region.
A part of the swept-up mass flows into the \ion{H}{II} region, and
the other part remains in the shell. The total swept-up mass 
at the time $t$ is written as, 
\begin{eqnarray}
M_{\rm sw}(t) &=& \frac{4 \pi}{3} \rho_{\rm c} R_{\rm c}^3 
                + \int_{\rm R_{\rm c}}^{R(t)}
                    \rho_{\rm c}
                    \left( \frac{r}{R_{\rm c}} \right)^{-w}
                    4 \pi r^2~dr    \nonumber \\
              &\simeq& \frac{4 \pi}{3 - w}
                       \rho(R) R(t)^3 ,   
\label{eq:msw}
\end{eqnarray}
where $R(t)$ is the position of the SF, and $\rho(R)$ is the
initial mass density at $r = R$ given by equation (\ref{eq:nr}).
The mass of the ionized hydrogen in the \ion{H}{II} region is given by  
\begin{equation}
M_{\rm i}(t) = \frac{4 \pi}{3} \rho_{\rm i} R_{\rm i}(t)^3 ,
\label{eq:mi}
\end{equation}
where $R_{\rm i}(t)$ is the position of the IF, and
$\rho_{\rm i}(t)$ is the mass density in the \ion{H}{II} region at
the time, $t$. Since the total recombination rate within the \ion{H}{II} region
always balances the UV-photon number luminosity, the \ion{H}{II} density 
decreases as $\rho_{\rm i} \propto R_{\rm i}^{-3/2}$.
Presuming $R(t) \sim R_{\rm i}(t)$, equations (\ref{eq:msw})
and (\ref{eq:mi}) mean that the mass ratio, 
$M_{\rm sw}(t)/M_{\rm i}(t)$ is proportional to 
$\rho(R)/\rho_{\rm i}(t) \propto R^{(3 - 2w)/2}$.   
Since the SF appears at $R_{\rm i} \sim R_w$, we get
\begin{equation}
\frac{M_{\rm sw}(t)}{M_{\rm i}(t)}
\simeq \left[ \frac{R(t)}{R_w} \right]^{(3 - 2w)/2} .
\label{eq:fm}
\end{equation}
Using equations (\ref{eq:msw}), (\ref{eq:mi}) and (\ref{eq:fm}),
the \ion{H}{II} mass density is written as,
\begin{equation}
\rho_{\rm i}(t) \simeq \frac{3}{3-w} \rho(R)
                       \left( \frac{R_w}{R(t)} \right)^{(3-2w)/2} .
\label{eq:rhoi}
\end{equation}
In this paper, we focus on the structure of the outer PDR as well 
as the \ion{H}{II} region. The PDR structure depends on the efficiency of 
the shielding of FUV radiation from the central star.
Since the FUV radiation is attenuated by the dust absorption and 
self-shielding effect, the penetration of the PDR is sensitive
to the column density of the shell and outer envelope. 
First, let us evaluate the time evolution of the column density
of the shell. Calculating $M_{\rm sw}(t) - M_{\rm i}(t)$ with 
equations (\ref{eq:msw}), (\ref{eq:mi}), and (\ref{eq:rhoi}), the 
mass of the shell at the time $t$ is,
\begin{equation}
M_{\rm sh}(t) = \frac{4 \pi}{3 - w}
                \rho(R) R(t)^3 
             \left[ 1 - 
                       \left( \frac{R_w}{R(t)} \right)^{(3-2w)/2}
             \right] .
\end{equation}
Therefore, the column density of the shell is,
\begin{equation}
N_{\rm sh}(t) = \frac{n_{\rm c} R_{\rm c}}{3-w}
                \left( \frac{R(t)}{R_{\rm c}} \right)^{1-w}
                \left[
                 1 - \left( \frac{R_w}{R(t)} \right)^{(3-2w)/2}
                \right] .
\label{eq:nsh}
\end{equation}
Equation (\ref{eq:nsh}) shows an interesting feature of the
time evolution of the shell column density. 
As the \ion{H}{II} region expands, the shell column density increases
with $w < 1$, but decreases with $w > 1$.
The critical power-law index is 
$w = 1$, with which the shell column density converges to the
constant value of $n_{\rm c} R_{\rm c}/2$.
If the density gradient is as steep as $w \sim 1.5$, the column
density of the shell is significantly reduced.
Supposing the finite size of the cloud, the inner edge of 
the envelope is the IF and the outer edge is the end of the cloud.
The column density of this outer envelope is, 
\begin{eqnarray}
N_{\rm ev}(t) &=& \int^{R_{\rm out}}_{R(t)}
                  n(r)~dr
             \nonumber \\
              &=&
                \frac{n_{\rm c} R_{\rm c}}{w-1}
                \left[
                \left( \frac{R_{\rm c}}{R(t)} \right)^{w-1} -
                \left( \frac{R_{\rm c}}{R_{\rm out}} \right)^{w-1}
                \right] 
\label{eq:nev}
\end{eqnarray}
for $w \neq 1$, where $R_{\rm out}$ is the radius of the outer 
limit of the cloud.
Equation (\ref{eq:nev}) shows that the column density of the 
envelope is mainly determined by $R_{\rm out}$
with $w<1$, but $R(t)$ with $w>1$.
Therefore, if the density gradient is as steep as $w > 1$, 
$N_{\rm ev}$ quickly decreases as the \ion{H}{II} region expands.
Equations (\ref{eq:nsh}) and (\ref{eq:nev}) predict that the
time evolution of the outer PDR qualitatively changes 
across $w \sim 1$. 
The quantitative changes will be significant, 
when the initial column density of the core is about
$N_{\rm H} \sim 4-8 \times 10^{21}~{\rm cm}^{-2}$.
In a molecular cloud stratified with $w < 1$, 
the column density of the shell increases as the 
\ion{H}{II} region expands, and the dust absorption becomes significant 
in the shell. 
The molecules are shielded from the FUV radiation by the dust, 
and will accumulate in the shell.
This is similar to the evolution in the homogeneous ambient medium
studied in Papers I and II.
With $w > 1$, on the other hand, the shell column density finally 
decreases as the \ion{H}{II} region swells up.
The dust absorption does not work in the shell, and the PDR 
extends far beyond the shell. Furthermore, equation (\ref{eq:nev})
shows that the column density of the outer envelope quickly decreases
as the \ion{H}{II} region expands. Therefore, it becomes easier and easier
for the FUV photons to penetrate and photodissociate the outer molecular 
envelope.  In the following sections, we verify our predictions, performing
detailed numerical calculations.

%%%%%%%%%%%%%%%%%%%%%%%%%%%%%%%%%
\section{Numerical Calculations}
\label{sec:calc}
%%%%%%%%%%%%%%%%%%%%%%%%%%%%%%%%%

%-------------------------------------------------------------%
\begin{figure}
\resizebox{\hsize}{!}{\includegraphics{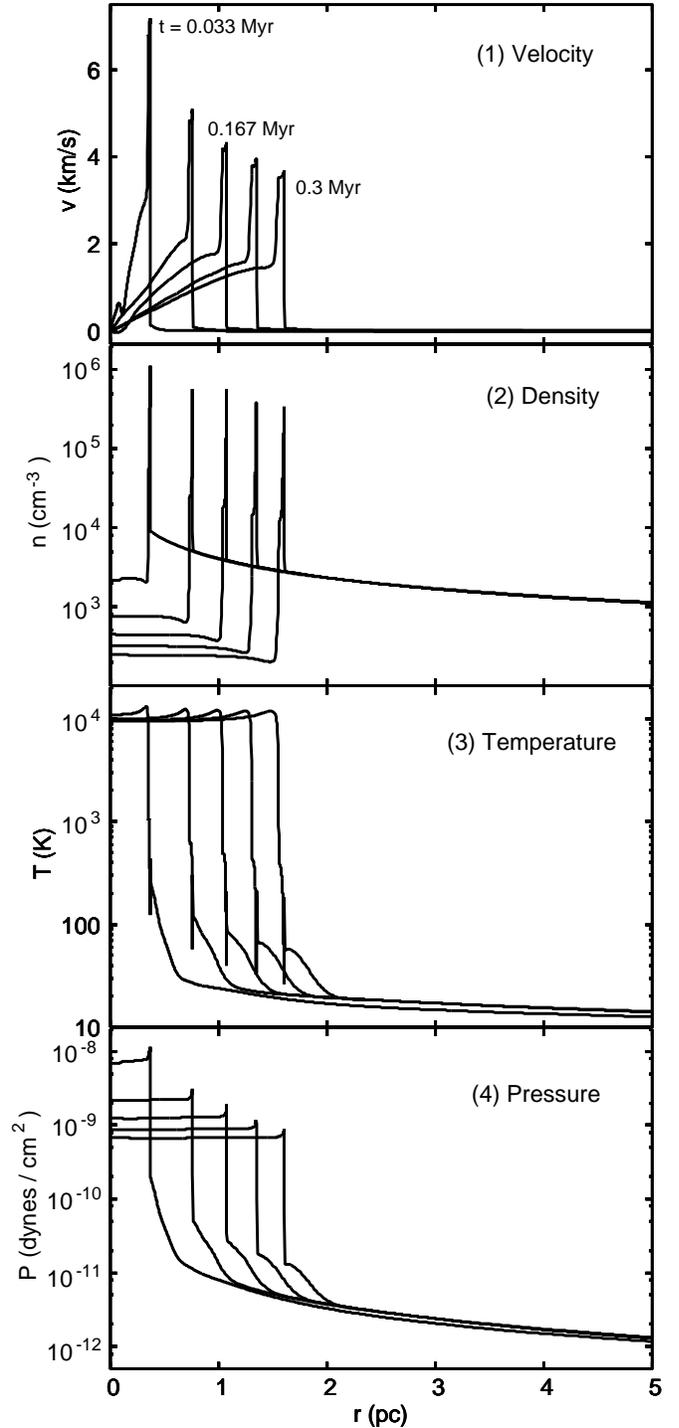}}
\caption{Gas-dynamical evolution with the density power-law
index, $w = 0.8$. In each panel, we present five snapshots
at $t=0.033$, 0.1, 0.167, 0.23, and 0.3~Myr.
}
\label{fig:hev_w08}
\end{figure}

\begin{figure}
\resizebox{\hsize}{!}{\includegraphics{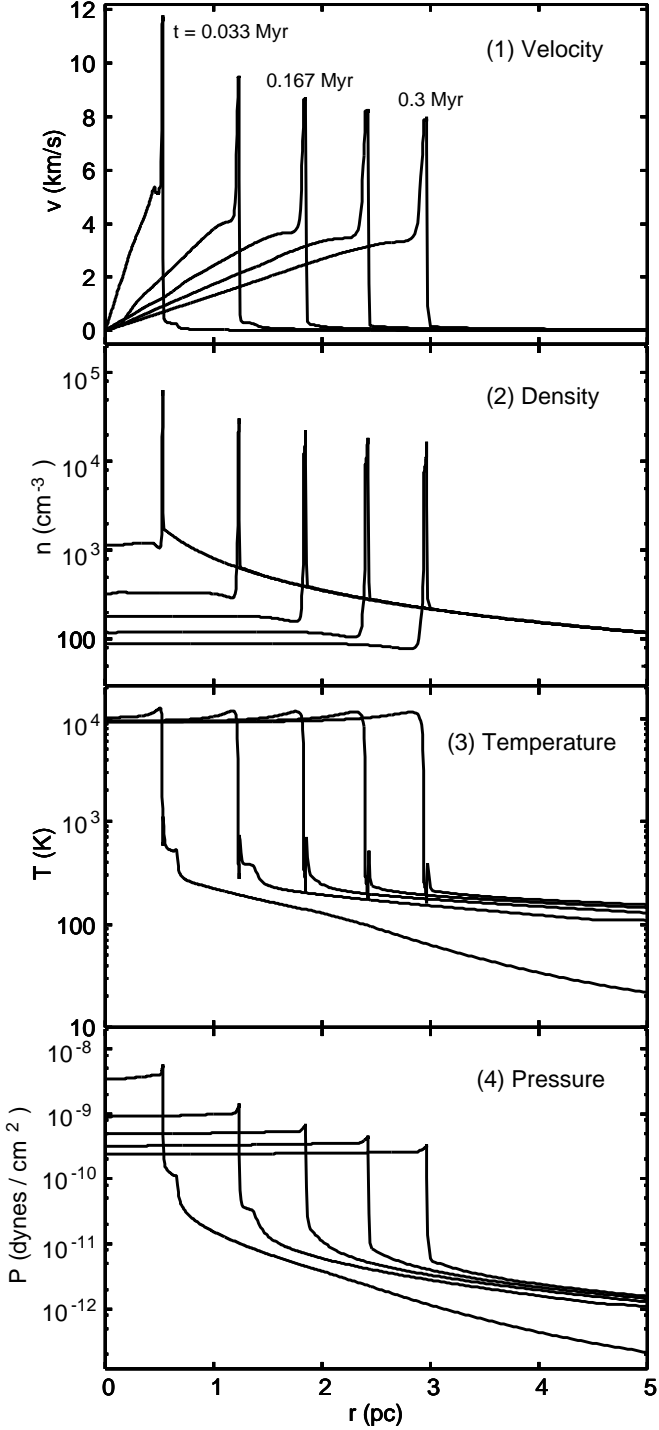}}
\caption{Same as Fig.\ref{fig:hev_w08} but for the 
density profile with $w = 1.2$.}
\label{fig:hev_w12}
\end{figure}
%--------------------------------------------------------------%

In this section, we present our numerical results calculated with our
code developed in Papers I and II (see Paper II for detailed
description), with which we can solve the time evolution 
of the thermal and chemical structure of the outer PDR as well as of the 
dynamical evolution of the \ion{H}{II} region. 
The UV/FUV radiative transfer and some chemical rate equations
are consistently solved with the hydrodynamics.
The dust absorption is included only outside the HII region.
We adopt the ambient density distribution given by equation 
(\ref{eq:nr}) with $n_{\rm c} = 10^5~{\rm cm}^{-3}$, and
the exciting star of $40.9~M_\odot$.
The corresponding stellar UV- and FUV-photon number luminosities are 
$S_{\rm UV} = 6.0 \times 10^{48}~{\rm s}^{-1}$ and
$S_{\rm FUV} = 5.8 \times 10^{48}~{\rm s}^{-1}$ \citep{DFS98}.
The Str\"omgren radius with the adopted $n_{\rm c}$ and
$S_{\rm UV}$ is $R_{\rm st} \sim 2.5 \times 10^{-2}$~pc, and
we set the core radius as $R_{\rm c} = 0.7 R_{\rm st} \sim
1.75 \times 10^{-2}$~pc. 
The IF and the DFs are initially set at 
$r = 0.2~R_{\rm c}$ and $r = 0.4~R_{\rm c}$ respectively.

Figures \ref{fig:hev_w08} and \ref{fig:hev_w12} present the 
gas-dynamical evolution with $w = 0.8$ and $1.2$ respectively. 
The dynamical features are similar in both models. 
When the IF reaches the radius, $R_w$, the SF appears in front of
the IF. FTB90 have provided the approximate equation 
for the time evolution of the \ion{H}{II} radius in this phase,
\begin{equation}
R(t) \simeq R_w
            \left[
             1 + \frac{7 - 2w}{4}
             \sqrt{ \frac{12}{9 - 4w} }
             \frac{t}{t_{\rm dyn}}
            \right]^{4/(7-2w)} , 
\label{eq:rana}
\end{equation}
where $t_{\rm dyn}$ is defined as $R_w/C_{\rm II}$, and
$C_{\rm II}$ is the sound speed in the \ion{H}{II} region 
($T \sim 10^4$~K).
Equation (\ref{eq:rana}) still provides a good approximate
evolution for our numerical calculations.
In both models, the density of the shell is 10-100
times as dense as that just ahead the SF. 
The expansion is faster with the steeper density gradient.
The size of the \ion{H}{II} region with $w=1.2$ is about twice as large
as that with $w=0.8$ at the same time, $t$.
The outward velocity within the \ion{H}{II} region is larger with the
steeper density gradient.

%---------------------------------------------------------------------------%
\begin{figure}
\resizebox{\hsize}{!}{\includegraphics{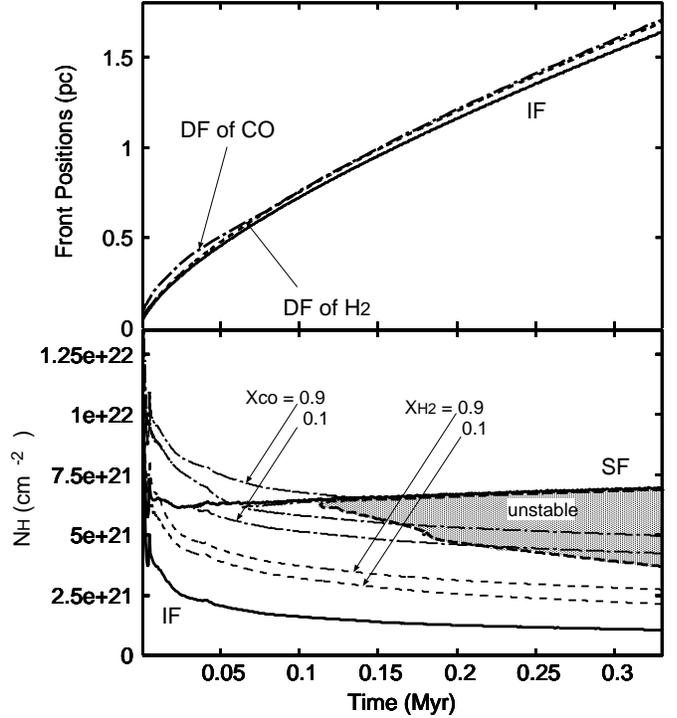}}
\caption{{\it Upper panel :} Time evolution of some front positions
in the model with $w = 0.8$.
The solid, broken, and dot-solid lines represent the positions of
the IF, DF of the H$_2$ and CO molecule at each time step.
We do not present the position of the SF, which almost traces
the position of H$_2$ DF.
{\it Lower panel :} The time evolution of the column density
of each region in the same model. The upper (lower) solid
line indicates the position of the SF (IF). Two broken (dot-solid)
lines show the position where $x_{\rm H_2} 
(x_{\rm CO})= 0.1$ and 0.9.  
The shaded region denotes the expected gravitationally unstable
region, where $t > 1/\sqrt{G \rho}$.
}
\label{fig:fev_w08}
\end{figure}

\begin{figure}
\resizebox{\hsize}{!}{\includegraphics{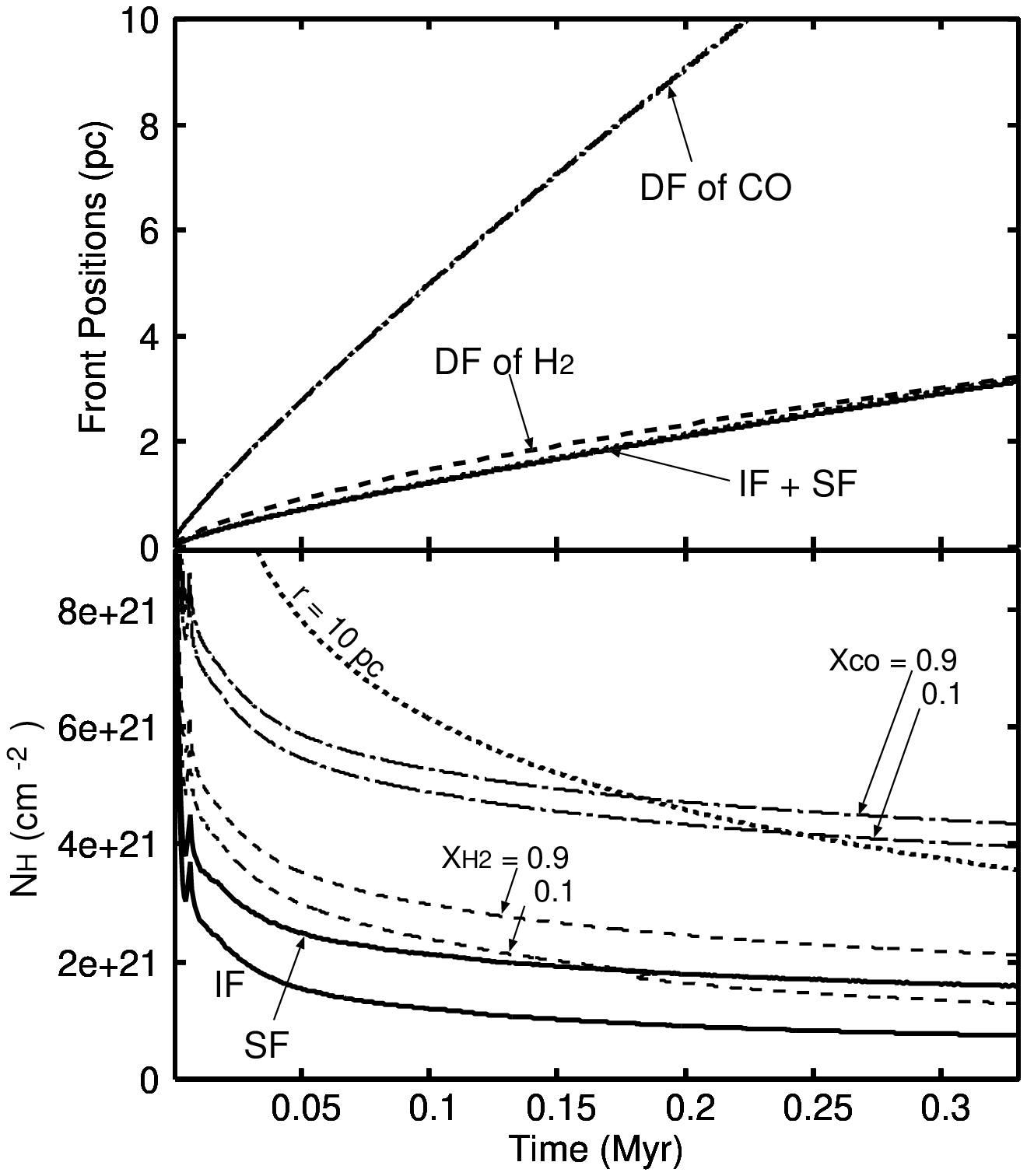}}
\caption{ Same as Fig.\ref{fig:fev_w08} but for $w = 1.2$.
We plot the position of the SF with a dotted line in the upper panel.
Note the differences in scale of the vertical axis from 
Fig.\ref{fig:fev_w08}.
The dotted line in the lower panel represents the column density
at $r \sim 10$~pc. Almost all CO molecules within 10~pc are photodissociated
by the time of $t \sim 0.2$~Myr.
}
\label{fig:fev_w12}
\end{figure}
%---------------------------------------------------------------------------%

\begin{figure}
\resizebox{\hsize}{!}{\includegraphics{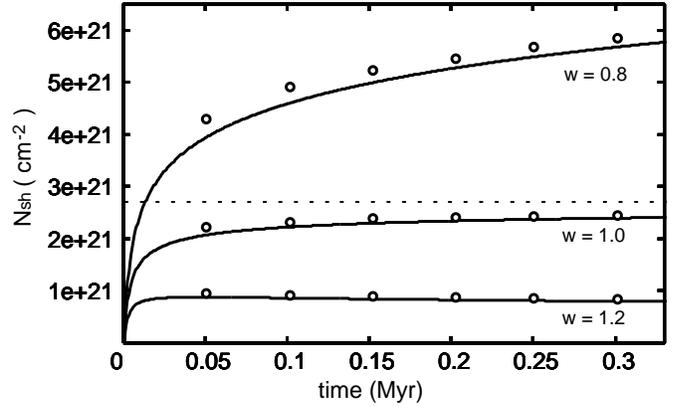}}
\caption{Time evolution of the column density of the shell
with the different density distributions. 
The three solid lines represent the time evolution expected by
equations (\ref{eq:nsh}) and (\ref{eq:rana}) with $w = 0.8$, 1.0 and 1.2 
from top to bottom. 
The open circles denotes the shell column density calculated 
every 0.05~Myr in our numerical models.
The horizontal dashed line indicates the half 
column density of the initial homogeneous core,  
$n_{\rm c}R_{\rm c}/2 \sim 2.7 \times 10^{21}~{\rm cm}^{-2}$.}
\label{fig:comp}
\end{figure}

Despite their similar dynamical evolution, the thermal and chemical 
structures of these model's shells and outer envelopes are 
very different.
The third panels of Figures \ref{fig:hev_w08} and \ref{fig:hev_w12}
show the temperature profile in each model. 
With the density gradient of $w=0.8$ (Fig.\ref{fig:hev_w08}), 
the temperature in the outer envelope is kept at $T \sim 10 - 30$~K,
except for just outside the shell. 
With the steeper density gradient of $w=1.2$ (Fig.\ref{fig:hev_w12}),
however, the outer envelope is heated up to $T \sim 100 - 300$~K in the
initial 0.1~Myr.
Figure \ref{fig:fev_w08} shows the chemical structure of the
shell and outer envelope with $w = 0.8$ at each time step. 
The DFs of both H$_2$ and CO molecules are trapped in the shell 
by the time of $t \sim 0.1$~Myr. The molecular gas gradually 
accumulates in the outer region of the shell. The outer envelope 
surrounding the \ion{H}{II} region remains molecular. 
In the model with $w = 1.2$, on the other hand, 
we can see the different chemical evolution in the upper panel of 
Figure \ref{fig:fev_w12}. While the H$_2$ DF remains just in 
front of the SF, the CO DF leaves the SF and quickly travels over 
the whole cloud. Almost all CO molecules within 10~pc are 
photodissociated by the time of $t \sim 0.2$~Myr.
These clear differences are due to the different evolution
of the column densities of the shell and outer envelope, which
was predicted in \S~\ref{sec:ana}.
In the model with $w=0.8$, the column density of the shell
increases as the \ion{H}{II} region expands, as the lower panel
of Figure \ref{fig:fev_w08} and Figure \ref{fig:comp} show.
The shell column density attains $N_{\rm sh} \sim 5 \times
10^{21}~{\rm cm}^{-2}$, which corresponds to A$_{\rm V} \sim 2.5$, 
at $t \sim 0.1$~Myr.
The significant dust absorption in the shell efficiently blocks
FUV photons from the central star.
The FUV radiation field in Habing units \citep{Hb68} is 
$G_{\rm FUV} \sim 10^4$ at the IF, but only $\sim 50$ at the SF at
$t \sim 0.1$~Myr. Consequently, the gas temperature of the outer
envelope is as low as $T \sim 10-30$~K, owing to the 
weak FUV radiation field.
Both H$_2$ and CO molecules are protected against FUV photons
by the dust absorption and self-shielding effect, and DFs are
easily trapped in the shell.
After the DF is engulfed in the shell, the SF sweeps up the 
molecular gas in the envelope. The molecular gas accumulates in 
the shell; this is basically the same time evolution as in a 
homogeneous ambient medium (see Papers I and II).

%------------------------------------------------------------------%
%--------------------------------------------------------------------%

With the steeper density gradient with $w=1.2$, our model
shows the different time evolution of the column density in 
each region 
(the lower panel of Fig.\ref{fig:fev_w12} and Fig.\ref{fig:comp}).
As expected with equation (\ref{eq:nsh}), the column density
of the shell decreases as the \ion{H}{II} region expands.
Since the shell column density never exceeds 
$1.0 \times 10^{21}~{\rm cm}^{-3}$, the dust absorption in the
shell is not sufficient to protect molecules.
The self-shielding effect barely enables the gradual accumulation
of H$_2$ molecules in the shell, but the FUV radiation easily
photodissociates all CO molecules in the shell.
Furthermore, the column density of the outer envelope also
quickly decreases following equation (\ref{eq:nev}).
The column density becomes too low to shield the FUV radiation from 
the central star by the dust absorption in the envelope.
For example, the column density at $r \sim 10$~pc is initially
higher than $2.0 \times 10^{22}~{\rm cm}^{-3}$, but lower
than $4.0 \times 10^{21}~{\rm cm}^{-3}$ at $t \sim 3$~Myr 
(Fig.\ref{fig:fev_w12}). 
The DF of the CO molecule leaves the SF, and travels over the
whole molecular cloud. 

In the cloud with a much steeper density gradient, it becomes much
easier for the DFs to escape from the dense central region of the cloud.
As referred to in \S~\ref{sec:ana}, the \ion{H}{II} region can be trapped in 
the cloud with a density stratification of $w \leq 1.5$ (FTB90).
Therefore, only the PDR can broadly extend around the trapped 
\ion{H}{II} region with the density profile of $1 \leq w \leq 1.5$.
With such a steep density gradient, and if
the DFs reach the diffuse outskirts of the cloud, 
a density-bounded PDR should be observed around the 
photon-bounded \ion{H}{II} region.

%%%%%%%%%%%%%%%%%%%%%%%%%%%%%%%%%%%%%%%%%%%%%%%%%%%%%%%%%%%%%%%%%%%
\section{Application to the Galactic \ion{H}{II} Region ; Sharpless 219}
\label{sec:sh219}
%%%%%%%%%%%%%%%%%%%%%%%%%%%%%%%%%%%%%%%%%%%%%%%%%%%%%%%%%%%%%%%%%%%

%---------------------------------------------------------------------%
\begin{figure}
\resizebox{\hsize}{!}{\includegraphics{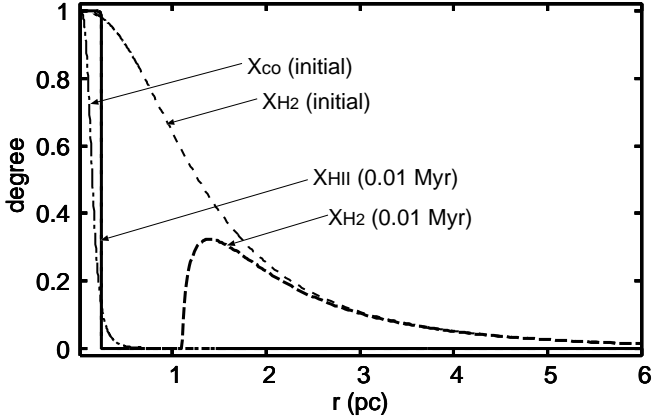}}
\caption{Snapshots of the chemical structure of our model for Sh219.
The thin dot-solid and broken line represent the initial molecular ratios,
$X_{\rm CO} \equiv n_{\rm CO}/n_{\rm H, nuc} Z_{\rm C}$ and 
$X_{\rm H_2} \equiv 2 n_{\rm H_2}/n_{\rm H,nuc}$, where
$n_{\rm H, nuc}$ is the number density of the hydrogen nuclei.
The thick solid and broken line represent the ratios,
$X_{\rm HII} = n_{\rm H^+}/n_{\rm H,nuc}$ and $X_{\rm H_2}$ at
$t = 0.01$~Myr. 
}
\label{fig:x_sh219}
\end{figure}
%---------------------------------------------------------------------%

%---------------------------------------------------------------------%
\begin{figure}
\resizebox{\hsize}{!}{\includegraphics{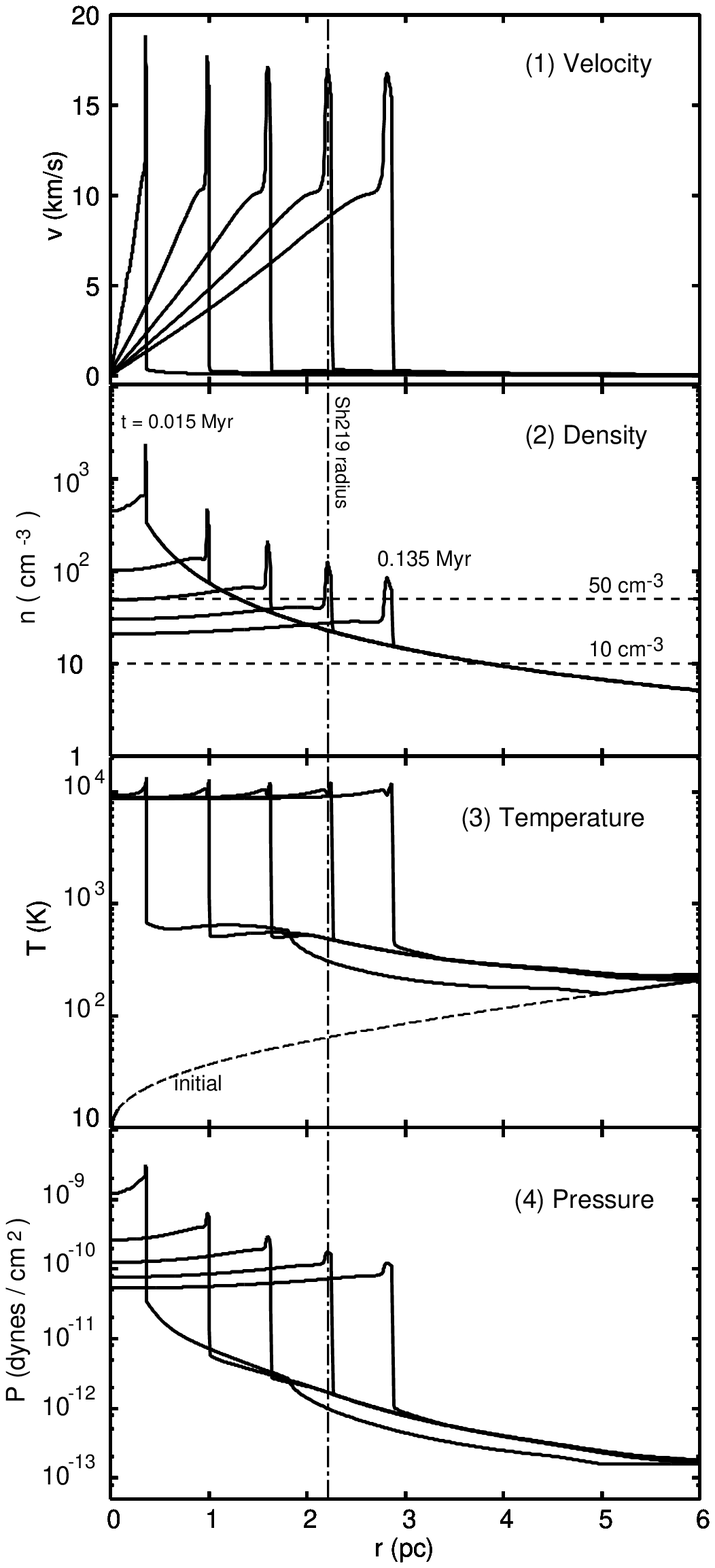}}
\caption{Gas-dynamical evolution of our model for Sh219.
The adopted power-law index of the density profile is $w=1.5$.
We present five snapshots at the time of $t = 0.015$, 0.045,
0.075, 0.105, and 0.135~Myr in each panel. The vertical dot-solid line
indicates the observed radius of Sh219, $\sim 2.2$~pc. }
\label{fig:hev_sh219}
\end{figure}
%---------------------------------------------------------------------%

\subsection{A Model with $w = 1.5$}

In this section, we focus on the Galactic \ion{H}{II} region, Sharpless 219
(Sh219). Sh219 shows a spherical morphology around a single B0V star;
its radius is about 2.2~pc. As mentioned above, Sh219 shows some
properties different from those of the ``collect and collapse'' 
\ion{H}{II} regions.
For example, the expanding shell around the \ion{H}{II} region has not
detected by the molecular emission.  Only the half-ring of the
PAH emission surrounds the ionized gas \citep{Deh03a, Deh05}.
Instead of a dense molecular shell, a diffuse atomic layer is
widely distributed around the \ion{H}{II} region, and the diffuse far-IR 
emission has been detected in this region \citep{RL93}. 
The radial width of the neutral layer is $2 - 3$~pc. 
The averaged number density of the layer is
only $\sim 9~{\rm cm}^{-3}$, which is as low as the density typical 
of the cold neutral medium, rather than that of the molecular cloud.
The number density in the \ion{H}{II} region has been
estimated as $55~{\rm cm}^{-3}$ \citep{RL93} 
and $170~{\rm cm}^{-3}$ \citep{Deh00}.  
Therefore, the ionized gas is much denser than the surrounding neutral
layer. These features sharply contrast with those of a ``collect and
collapse'' \ion{H}{II} region, for example, Sharpless 104 
\citep[Sh104;][]{Deh03b}.
Sh104 is also a pc-scale spherical \ion{H}{II} region, but surrounded
by a dense molecular shell. The estimated average density and molecular
mass of the shell are about $6000~{\rm cm}^{-3}$ and $6000~M_\odot$
respectively. The number density in the \ion{H}{II} region is 
$\sim 80~{\rm cm}^{-3}$, which is much lower than that of the shell.
We have shown that these observational properties of Sh104 are well
explained by our model, where the \ion{H}{II} region swells in the homogeneous
ambient medium of $\sim 10^3~{\rm cm}^{-3}$ (Paper I).
In this subsection, we propose an alternative model for Sh219, where the 
\ion{H}{II} region is expanding in a radially stratified molecular cloud.

We adopt the initial density distribution given by equation (\ref{eq:nr})
with $w = 1.5$. We adopt a $19~M_\odot$ star as the central star,
and the corresponding UV and FUV luminosities are 
$S_{\rm UV} = 5.6 \times 10^{47}~{\rm s}^{-1}$ and
$S_{\rm FUV} = 1.6 \times 10^{48}~{\rm s}^{-1}$ \citep{DFS98}.
The time evolution quantitatively changes with the different
parameters of the density and radius of the core. In this section, 
we present a model with $n_{\rm c} = 10^5~{\rm cm}^{-3}$ and 
$R_{\rm c} = 0.7~R_{\rm st}  \sim 0.07$~pc. 
The observed neutral layer around Sh219 is so diffuse
that we take into account the fact that the FUV background radiation (FBR)
penetrates into the cloud. The initial thermal and chemical
structure of the cloud are set at the equilibrium determined
with the local number density and FUV radiation field.
We adopt $G_{\rm FUV} = 1.0$ in Habing units as the 
fiducial FBR. 
Note that the strength of the FBR does not affect the chemical/thermal
structure outside the \ion{H}{II} region when the IF reaches the observed
radius of Sh219 (see below).
The third panel of Figure \ref{fig:hev_sh219} shows that the gas temperature 
in the outer envelope is initially as high as  
100-200~K, owing to the photoelectric heating by the FBR. 
The photodissociation by the FBR is significant 
in the outer diffuse region. More than 90\% of H$_2$ (CO) molecules
is initially photodissociated at $r > 3$~pc ($r > 0.3$~pc)
(see Fig.\ref{fig:x_sh219}).
Figure \ref{fig:hev_sh219} shows the gas-dynamical evolution 
of our model for Sh219. The basic evolution is similar to the
models presented in \S~\ref{sec:calc}.
The IF and preceding SF expand, and reach the radius of Sh219,
$r \sim 2.2$~pc at $t \sim 0.1$~Myr.
At that time, the number density in the \ion{H}{II} region is 
$\sim 50~{\rm cm}^{-3}$, and that of the outer layer is
$\sim 10~{\rm cm}^{-3}$ at $r \sim 4$~pc. 
Therefore, the observed density structure of the Sh219 is naturally 
explained by our model.

%----------------------------------------------------------------------%
\begin{figure}
\resizebox{\hsize}{!}{\includegraphics{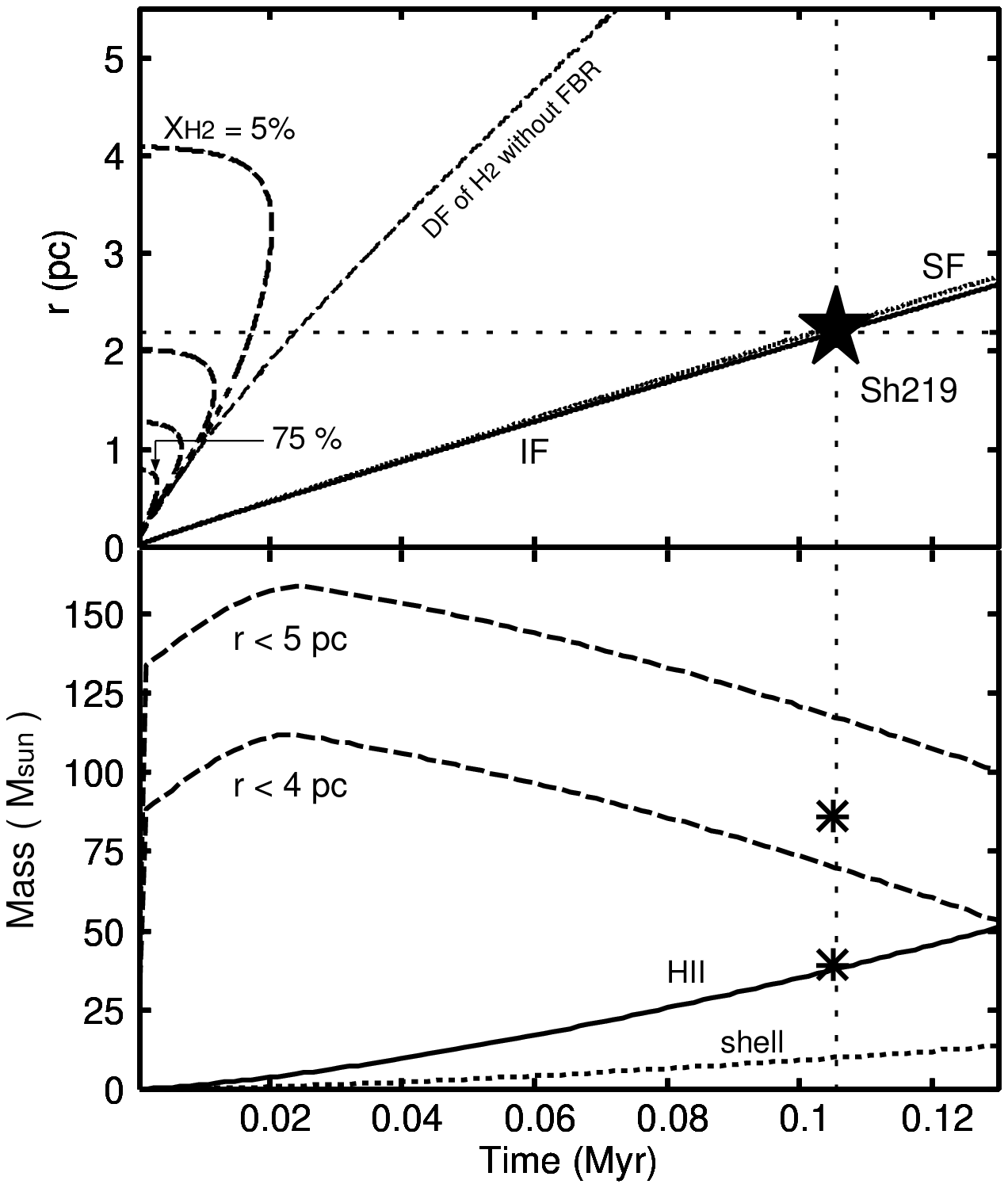}}
\caption{{\it Upper panel}: Time evolution of the positions of
some fronts in our model for Sh219. 
The solid and dotted line represent the positions
of IF and SF. The broken contours denote the positions where
the H$_2$ molecular percentages are 5\%, 25\%, 50\%, and 75\%
from up to down. The thin broken line shows the position 
of the H$_2$ DF without the FUV background radiation.  
The horizontal dashed line indicates the observed radius of
Sh219, $\sim 2.2$~pc. The \ion{H}{II} radius reaches the observed
radius at $t \sim 0.105$~Myr, which is indicated with vertical
dashed line.
{\it Lower panel}: Time evolution of the mass in each region.
The solid and dotted lines represent the mass of the ionized
hydrogen in the \ion{H}{II} region, and neutral hydrogen in the shell
respectively. Two broken lines show the mass of the diffuse neutral
layer in $r < 4$ and 5~pc. Asterisks indicate the observed mass 
of the \ion{H}{II} region ($34~M_\odot$)
and surrounding neutral layer ($86~M_\odot$) by \citet{RL93}. 
}
\label{fig:fev_sh219}
\end{figure}
%----------------------------------------------------------------------%

The molecular gas in the central dense region is quickly destroyed
by the FUV radiation from the central star.
H$_2$ molecules in the inner 1~pc are photodissociated over the
initial 0.01~Myr (see Fig.\ref{fig:x_sh219}).
The upper panel of Figure \ref{fig:fev_sh219} presents the time
evolution of the chemical structure of the envelope.
As this panel shows, H$_2$ molecules in the envelope completely
disappear by the estimated age of Sh219, $t \sim 0.1$~Myr. 
Almost all H$_2$ molecules are dissociated by  
$\sim 0.04$~Myr, and CO molecules in the inner region are 
destroyed just after the start of the calculation. 
The \ion{H}{II} region is surrounded with the diffuse atomic layer, where the
gas temperature is about $300 - 500$~K.
Our model suggests that the observed wide neutral layer around Sh219
should not be photon-bounded, but density-bounded.
This does not depend on the strength of the FBR.
In Figure \ref{fig:fev_sh219}, we also plot the propagation of
the H$_2$ DF when the FBR is not included. 
Even without the FBR, FUV photons from the central star
easily penetrate into the molecular envelope, and are sufficient
for the photodissociation of almost all H$_2$ molecules in the envelope. 
This is because the column densities of the shell and envelope
are too low to block FUV photons
with the steep density gradient of $w = 1.5$ 
(see \S~\ref{sec:ana}).

Figure \ref{fig:fev_sh219} shows the time evolution of the mass 
of the \ion{H}{II} region and outer neutral layer. 
Assuming that the neutral layer is density-bounded at 
$r \sim 4-5$~pc, which is the observed outer radius of the H~I layer, 
the calculated masses at $t \sim 0.1$~Myr, when the IF reaches the
observed radius of Sh219, show good agreement with the observed values.
The mass of the H~I layer, $M_{\rm HI}$, initially increases, 
because the H$_2$ molecules in the envelope are photodissociated 
by the FUV photons from the central star. 
After all H$_2$ molecules are destroyed ($t \sim 0.02$~Myr), $M_{\rm HI}$
decreases, because the \ion{H}{II} region gradually erodes
the H~I layer.
Note that the mass of the shell is much smaller than that of the
\ion{H}{II} region and neutral layer.
Most of the swept-up gas does not remain in the shell, but flows
into the \ion{H}{II} region.

\subsection{Timescale of Induced Star Formation}

Although Sh219 does not have a clear molecular shell,
some observations have suggested that star formation is 
actually triggered in a molecular cloud present at the periphery of Sh219 
\citep{Deh05, Deh06}. A young stellar cluster is embedded in the 
molecular cloud, and elongated along the IF.
The cluster includes several massive 
stars exciting a ultra-compact \ion{H}{II} region. 
One possible triggering scenario is that the IF and the 
preceding SF enter the pre-existing molecular cloud, and the 
dense compressed layer forms only in the cloud. The fragmentation of the 
compressed layer occurs and triggers star formation. 
Here, we examine this scenario, evaluating the timescale of this 
triggering process.
 
The average number density of the molecular cloud is estimated as 
$n({\rm H_2}) = 8.0 \times 10^3~{\rm cm}^{-3}$ \citep{Deh06}.
In our model, the expanding velocity of the IF 
and SF is $2~c_{\rm HII} \sim 20$~km/s 
($c_{\rm HII}$ is the sound speed in the HII region). 
After the IF and SF hit the molecular cloud, the SF can still 
propagate at $\sim$ several - 10 km/s within the photo-evaporating 
cloud \citep{Bd89}. If the cloud size is much smaller than
the \ion{H}{II} radius, the SF quickly transverses and compresses 
the cloud. The Mach number of the SF will be of the order of 10, 
and the density jump at the isothermal SF is, 
\begin{equation}
\frac{n_2}{n_1} \sim {\cal M}^2 ,
\end{equation}
where $n_1$ and $n_2$ are number densities ahead/behind the SF. 
With the average cloud number density, the number density 
just behind the SF can be $n({\rm H_2}) \sim 10^6~{\rm cm}^{-3}$. 
The timescale of the fragmentation is,
\begin{equation}
t_{\rm frg} \sim \frac{1}{\sqrt{G \rho}} 
            \sim 8.4 \times 10^4~{\rm yr}
            \left(
             \frac{n({\rm H_2})}{10^6~{\rm cm}^{-3}}
            \right)^{-1/2} ,
\end{equation}
and $t_{\rm frg}$ with $n({\rm H_2}) \sim 10^6~{\rm cm}^{-3}$ is
 comparable with the estimated age of Sh219, $\sim 0.1$~Myr. 
Therefore, triggering is possible.
This is also comparable with the typical age of the ultra-compact
\ion{H}{II} regions \citep[e.g.,][]{WC89, Ch02}.
Note that if Sh219 is modeled in a homogeneous ambient
medium of $30-50~{\rm cm}^{-3}$, the estimated age is much younger,
only a few times $10^4$~yr \citep{RL93}. In this case,
some special conditions should be satisfied for the rapid 
triggering scenario: The SF must enter the region much denser
than average density, and subsequent star (cluster) formation
must advance very rapidly.

%%%%%%%%%%%%%%%%%%%%%
\section{Discussions}
\label{sec:dis}
%%%%%%%%%%%%%%%%%%%%%

\subsection{Implication for the Feedback Effect 
            : Negative or Positive ?}

In this subsection, we discuss the role of UV/FUV radiation
for star formation in the molecular cloud.
\citet{RD92} and \citet{DFS98} have calculated the expansion of 
the \ion{H}{II} region and PDR in a homogeneous ambient medium,  
solving the UV/FUV radiative transfer around a massive star.
They have shown that a significant amount of the molecular gas
is photodissociated by the FUV radiation.
\citet{DFS98} have suggested that the star formation efficiency
of the molecular cloud is significantly suppressed by the 
photodissociation ({\it negative feedback}).
However, our up-to-date calculations, which solve the hydrodynamics
as well as the radiative transfer, have shown another
aspect of the feedback effect (Papers I and II).
We have shown that the dense shell forms, and that most of the
swept-up gas remains in the shell in a homogeneous medium. 
While the FUV radiation destroys the molecular material, FUV photons
are finally shielded by the high column density of the shell. 
The dense molecular shell forms, where triggering of the
subsequent star formation occurs ({\it positive feedback}). 
Our quantitative analysis has shown that the positive 
feedback effect can dominate the negative effect in some cases with
the homogeneous medium \citep{HI06b}. 

Our current work has shown that the negative feedback effect
is promoted only in the inhomogeneous medium.
In the cloud with a steep density gradient of $w>1$, the shielding
in the shell and envelope becomes inefficient as the \ion{H}{II}
region expands. The PDR can widely extend around the ionized gas.
Consequently, the star formation efficiency will be 
reduced, though triggering still can occur in the adjacent
molecular clouds, as shown in Sh219.
Ultimately, the feedback effect by the UV/FUV radiation should be 
examined in the clumpy, turbulent medium. 
Some recent studies have calculated the expansion of the \ion{H}{II}
region in this realistic situation \citep[e.g.,][]{Ml06, McL06}, and 
suggested that the clumpy structure may also diminish the positive 
effect \citep{Dl05}.  Although these efforts have not included the outer PDR, 
similar approaches solving the FUV radiative transfer will make it possible
to clarify the role of the FUV radiation.
For this purpose, however, the swept-up shell should be resolved with
a sufficient number of grids. The column density of the shell often 
dominates the total column density of the system, which is important 
for the penetration of FUV photons.

%%%%%%%%%%%%%%%%%%%%%
\section{Conclusions}
\label{sec:conc}
%%%%%%%%%%%%%%%%%%%%%

In this paper, we have studied the time evolution of 
the \ion{H}{II} region and surrounding PDR in the radially stratified molecular
cloud. We have examined the efficiency of the trapping of 
FUV photons in clouds with different density profiles
represented as $n(r) \propto r^{-w}$.
We have focused on the expansion with $w \leq 1.5$.
In the molecular cloud with $w > 1.5$, neither the IF nor DFs are 
trapped, and quickly travel over the whole cloud.

The key physical quantity for the trapping of the FUV radiation
is the column density, because the dust absorption is the primary
shielding agent. 
First, we have analytically shown that the time evolution of the column 
densities of the shell and outer envelope qualitatively switches 
across $w=1$. The column density of the shell increases
as the \ion{H}{II} region expands in the cloud with $w < 1$.
With $w > 1$, however, the shell column density finally
decreases, and the column density of the outer envelope quickly
decreases as the \ion{H}{II} region expands. 
The quantitative difference across $w = 1$ is
significant when the initial column density of the core is 
$\sim {\rm several} \times 10^{21}~{\rm cm}^{-2}$. 

Next, we have verified the analytic consideration using the numerical
calculations. The chemical/thermal structure outside the \ion{H}{II} region
sharply changes with $w$. In the cloud with a steep density gradient
of $w > 1$, the PDR can extend broadly around the \ion{H}{II} region.
The stellar FUV radiation heats up the envelope to several $\times$
100~K via the photoelectric heating. The trapping of the DFs is
sometimes selective. Only CO DF can travel over the whole cloud, while
H$_2$ DF is trapped in the shell. 
In the steeper density gradient, the trapping of the DFs grows 
less efficient. If the density gradient is as steep as $w \sim 1.5$,
both H$_2$ and CO DFs quickly propagate in the cloud, and almost all 
molecules in the envelope are finally photodissociated.
This is contrasted with the expansion in the homogeneous medium
and the gradual density gradient of $w < 1$.
In these cases, DFs are finally engulfed by the shell, and the
molecular gas gradually accumulates in the shell (Papers I and II).

Finally, we have applied this case $(w = 1.5)$ to the 
Galactic \ion{H}{II} region, Sh219, whose observational properties 
are very different from the
``collect and collapse''candidates. In Sh219, the ionized gas is not 
surrounded by a dense molecular shell, but by a diffuse neutral layer.  
The \ion{H}{II} region is denser than the surrounding neutral layer.
Our model naturally explains these characters of Sh219.
The calculated size, density and mass of the \ion{H}{II} region
and neutral layer are in good agreement with the observed values
at $t \sim 0.1$~Myr.
We suggest that a density-bounded PDR surrounds the photon-bounded 
\ion{H}{II} region in Sh219.

\begin{acknowledgements}
I am grateful to Lise Deharveng for the useful comments
and careful reading of the manuscript. 
I also thank Shu-ichiro Inutsuka
for the continuous encouragement and fruitful discussions.
\end{acknowledgements}

\bibliographystyle{aa}
\end{document}